\documentclass[11pt]{article}
\usepackage[margin=1in]{geometry}
\usepackage[utf8]{inputenc}
\usepackage[T1]{fontenc}
\usepackage{lmodern}
\usepackage{microtype}
\usepackage{amsmath,amssymb}
\usepackage{booktabs}
\usepackage{graphicx}
\usepackage{caption}
\usepackage{subcaption}
\usepackage{xcolor}
\usepackage{listings}
\usepackage[hidelinks]{hyperref}
\usepackage{natbib}
\usepackage{authblk}
\usepackage{tabularx}

\newcommand{\gpt}{GPT-6 Astra}

\newcommand{\eg}{e.g.,}

\title{Weight-Space Programming of Variational Quantum Circuits:\\ Large Language Models Write Quantum Circuit Parameters Without an Optimizer}

\author[1]{Debarshi Kundu}
\author[1]{Swaroop Ghosh}
\author[1]{Vasant Honavar}
\affil[1]{The Pennsylvania State University}
\date{}

\begin{document}
\maketitle

\begin{abstract}
Variational quantum algorithms train the angles of a parameterized quantum circuit with a classical optimizer, and the optimizer is the bottleneck: every loss evaluation is a run of the quantum device, gradients vanish for generic circuits, and the problem is NP-hard in general. We ask whether a large language model can write the angles directly. \emph{Weight-space programming}, introduced by Takehi et al.\ for classical neural networks, has a general-purpose LLM read and write a network's weights directly, without an optimizer; we bring it to quantum circuits. The model is given the problem, the circuit and the gate conventions, and replies with the numbers; the circuit is simulated exactly as written, and no loss, gradient or accuracy is ever returned. We test a frontier model (\gpt) on the three main families of variational algorithms: classification of handwritten digits with a quantum circuit, ground-state energies of spin chains and H$_2$ (VQE), and MaxCut (QAOA), 139 requests in all, each compared against random weights, a classical optimizer, and the optimizer started from the model's weights on the same circuit. Where the weights have a meaning the model can reason about, it matches the optimizer: 95--100\% of the energy spectrum on every VQE problem, the exact H$_2$ ground state from a closed-form derivation, optimal depth-one QAOA angles on every graph, and 93--96\% accuracy on 3 vs 8 from 12--18 hand-written weights of a fixed circuit, above an optimizer trained on the same ten examples. Its weights are also a strong warm start: the optimizer started from them reaches its best value after a median of one evaluation instead of hundreds. Left to design the circuit, the model chooses, in every one of 30 cases, chains of data-dependent rotations whose weights are pixel importances (a single qubit for binary tasks, one such chain per qubit for ten classes), and programs them to 65\% accuracy on all ten digits. The method does not scale to opaque weight spaces (accuracy falls to chance as a fixed circuit is deepened past 30 weights while the optimizer improves) and is sensitive to conventions (6 of 42 QAOA replies negate the mixing angles and collapse). We release every prompt, reply and score.

\end{abstract}

\section{Introduction}
\label{sec:intro}

Every variational quantum algorithm has the same shape as a neural network: a fixed computational graph (the circuit) with continuous parameters (the gate angles), a loss computed by running the graph, and a classical optimizer that adjusts the parameters \citep{cerezo2021variational}. The optimizer is the expensive part. Each loss evaluation is a run of the quantum device, gradients vanish exponentially in the number of qubits for generic circuits \citep{mcclean2018barren}, and the optimization problem is NP-hard in general \citep{bittel2021training}. A great deal of work therefore goes into choosing good initial parameters, or into reusing parameters across problem instances.

This paper asks a simple question: \emph{can a large language model write the parameters directly?} Not propose a circuit for someone else to train, and not act as an optimizer that is shown scores and asked to improve them, but read a description of the problem---a Hamiltonian, a graph, ten pictures of handwritten digits---and reply with the numbers that go into the gates, with no feedback of any kind. \citet{takehi2026sparks} recently showed that a frontier model can do this for classical neural networks: read the weights of a trained network and say what it does, edit them to fix a failing policy, and write the weights of game-playing networks from scratch. They call it \emph{weight-space programming}---a general-purpose LLM reads and writes a network's weights directly, not through an optimizer, around a natural-language specification and sometimes with no environment feedback at all. Our work is inspired by theirs and carries the idea to a domain where it has a particular pull: variational quantum circuits have few parameters, each with a physical or geometric meaning, and the cost of the optimizer they replace is paid in runs of a quantum device.

We test the question on the three main families of variational quantum algorithms: supervised classification with a parameterized circuit (QML), ground-state energy estimation (VQE), and combinatorial optimization (QAOA). In every experiment the model receives the problem, the exact gate conventions and the circuit (or, in one QML setting, a small circuit language in which it may design its own), and returns a JSON list of angles. The angles are simulated exactly as written. We compare them against three references on the same circuit: random angles, a classical optimizer started from random angles, and the same optimizer started from the model's angles. All experiments use one frontier model, \gpt, and every prompt, reply and score is released.

\paragraph{Findings.}
\begin{enumerate}
  \item \textbf{Weight programming works when the weights have meaning.} With the Hamiltonian variational ansatz for the transverse-field Ising chain, the model's angles capture 95--100\% of the spectrum between the highest and lowest eigenvalue on every problem, equal to the optimizer; for H$_2$ it derives the energy as a function of the single angle and writes its exact minimizer; for QAOA at depth one it derives the closed-form expectation of the cut and maximizes it analytically, matching the optimizer on all six graphs. For digit classification it hand-writes 12--18 weights of a fixed amplitude-encoding circuit and scores 93--96\% on 3 vs 8, \emph{above} an optimizer trained on the same ten examples.
  \item \textbf{The model's weights are a strong warm start.} Started from the model's angles, the optimizer reaches its best value after a median of one loss evaluation in VQE and QAOA, against hundreds from random starts, and it reaches that value more often.
  \item \textbf{Left to design the circuit, the model picks the one it can reason about.} In the free-design QML setting the model converges, in 30 of 30 trials across six conditions, on chains of data-dependent rotations whose weights are pixel importances: a single-qubit chain for binary tasks (a linear classifier in disguise) and one such chain per qubit, read out compositionally, for ten classes, with hand-chosen weights that score 91--95\% on binary tasks and 65\% on all ten digits. When structural rules forbid the single-qubit chain, it finds circuit identities that compute the same weighted sum on two qubits, or writes a nearest-prototype classifier by hand. Interpretability of the weight space, not quantumness, is what the model exploits.
  \item \textbf{It does not scale to opaque weight spaces.} When the fixed circuit is deepened, so that each weight's effect on the output can no longer be reasoned about, accuracy on 3 vs 8 falls from 0.96 at 18 weights to 0.70 at 54, and on ten classes it never leaves chance (0.07--0.14) while the optimizer on the same examples climbs to 0.57. The failure is informative: the model says it is writing ``heuristic weights'' and it is right.
  \item \textbf{Conventions are a failure mode.} In 6 of 42 QAOA trials the model produced the correct angle magnitudes with every mixer angle negated; those circuits score near zero, and with the sign corrected they match the optimizer. Weight programming is only as good as the shared understanding of what a weight means.
\end{enumerate}

These results extend weight-space programming from classical networks to quantum circuits and reframe what a language model can contribute to variational quantum algorithms: not as a replacement for the optimizer in general, but as a source of parameters for circuits whose weights carry physical or geometric meaning, and as an initializer that removes most of the optimizer's work.

\section{Background for machine-learning readers}
\label{sec:background}

This section gives the minimum needed to read the rest of the paper. Readers familiar with variational quantum algorithms can skip to Section~\ref{sec:method}.

\paragraph{Qubits and states.}
A classical bit is $0$ or $1$. A \emph{qubit} is described by two complex numbers $(\alpha, \beta)$ with $|\alpha|^2 + |\beta|^2 = 1$; measuring it yields $0$ with probability $|\alpha|^2$ and $1$ with probability $|\beta|^2$. A register of $n$ qubits is described by $2^n$ complex numbers (the \emph{amplitudes}), one per bit string, with squared magnitudes summing to one. Measuring the register returns one bit string, drawn with probability equal to its squared amplitude. All the circuits in this paper are small enough ($\leq 8$ qubits) that we compute these $2^n$ amplitudes exactly on a laptop, so every number we report is a noise-free expectation, not a sampled estimate.

\paragraph{Gates are the layers; angles are the weights.}
A quantum circuit is a sequence of \emph{gates}, each a small unitary matrix applied to one, two, or three qubits. Some gates are fixed (the Hadamard gate $h$, the controlled-NOT $cx$), others are \emph{rotations} with a continuous angle: $ry(\theta)$ rotates one qubit by $\theta$ about the $Y$ axis of its state sphere, $rzz(\theta)$ couples two qubits. A circuit with free angles is a \emph{parameterized quantum circuit} (PQC). The analogy to a neural network is close and worth keeping in mind: the gate layout is the architecture, the angles are the weights, and the measurement is the readout. Two angles on the same qubit about the same axis compose by addition, $ry(a)\,ry(b) = ry(a+b)$, exactly as two consecutive linear layers without a nonlinearity collapse to one; entangling gates such as $cx$ play the role of the nonlinearity that keeps depth from collapsing.

\paragraph{Variational quantum algorithms.}
A variational quantum algorithm (VQA) \citep{cerezo2021variational} fixes a circuit layout, evaluates a loss by running the circuit and measuring, and updates the angles with a classical optimizer. The three families we study differ only in the loss:
\begin{itemize}
  \item \textbf{Quantum machine learning (QML).} The input $x$ is encoded into the circuit (as rotation angles proportional to pixel values, or as the amplitudes themselves), the angles are the trainable weights, and a measured probability is the class score \citep{mitarai2018quantum,schuld2020circuit,perezsalinas2020data}. The loss is a classification loss on labelled data.
  \item \textbf{Variational quantum eigensolver (VQE).} The loss is the energy $\langle \psi(\theta) | H | \psi(\theta) \rangle$ of the prepared state under a Hamiltonian $H$ (a Hermitian matrix describing a molecule or a magnet); the minimum over all states is the \emph{ground-state energy} $E_0$ \citep{peruzzo2014variational}. $H$ is written as a sum of Pauli terms, \eg\ $-\sum_i Z_i Z_{i+1} - h\sum_i X_i$ for a chain of spins with nearest-neighbour coupling and a transverse field $h$.
  \item \textbf{Quantum approximate optimization algorithm (QAOA).} The loss is the negative expected value of a combinatorial objective, here the MaxCut value of a graph \citep{farhi2014quantum}. The circuit alternates $p$ layers of a ``cost'' step $e^{-i\gamma_k C}$ and a ``mixing'' step $e^{-i\beta_k \sum_i X_i}$; the weights are the $2p$ angles $(\gamma_1, \beta_1, \ldots, \gamma_p, \beta_p)$.
\end{itemize}

\paragraph{Why the weights are the hard part.}
Training a VQA is, in general, NP-hard \citep{bittel2021training}, and generic circuits suffer from \emph{barren plateaus}: for random initializations the gradient of the loss vanishes exponentially in the number of qubits \citep{mcclean2018barren}, so an optimizer starting from random weights may learn nothing. Each loss evaluation is a run of the quantum device, so the number of evaluations an optimizer consumes is the dominant cost. Good initial weights are therefore valuable in their own right \citep{grant2019initialization,egger2021warm}, and for QAOA on regular graphs it is known that near-optimal angles are nearly instance-independent \citep{brandao2018fixed,zhou2020quantum,wurtz2021fixed}.

\paragraph{Two standard circuit layouts.}
The \emph{hardware-efficient ansatz} (HEA) \citep{kandala2017hardware} alternates a layer of single-qubit rotations with a ladder of $cx$ gates; it is generic and says nothing about the problem. The \emph{Hamiltonian variational ansatz} (HVA) \citep{wecker2015progress,wiersema2020exploring} builds layers from the terms of $H$ itself, so each angle has a physical meaning (how long to evolve under the coupling term, how long under the field term). We use both, because they differ in how much the structure of the problem is visible in the structure of the weights.

\section{Weight-space programming for quantum circuits}
\label{sec:method}

We follow the setting of \citet{takehi2026sparks}: a general-purpose model writes a network's weights directly, without an optimizer, from a natural-language specification. The network is a parameterized quantum circuit and the weights are its gate angles.

\paragraph{Protocol.}
A weight-space programming trial is a single request to a language model. The prompt contains (i) the problem---a Hamiltonian, a graph, or a handful of labelled images; (ii) the circuit, either fully specified with named angles $\theta[0..K{-}1]$, or a description of a small circuit language in which the model may design its own; (iii) the exact gate conventions (we use the Qiskit definitions, \eg\ $ry(t) = e^{-itY/2}$); and (iv) the reply format, a JSON object with a list of $K$ numbers (or a full circuit). The model's reply is simulated exactly as written. No loss value, gradient, or accuracy is ever sent back. The only feedback loop is syntactic: if the reply is not a valid circuit (wrong number of weights, a gate on a non-existent qubit, a malformed expression), the validation error is returned and the model may answer again, at most twice. Every prompt, reply, and parsed circuit is saved, and the model's free-text reasoning is kept alongside the numbers.

\paragraph{Two levels of freedom.}
In the \emph{weights-only} setting the circuit is fixed and the model writes $K$ real numbers. This is the clean test of the hypothesis, and it is the setting of all VQE and QAOA experiments and of the fixed-ansatz QML experiment. In the \emph{free-design} setting (QML only) the model also chooses the number of qubits, the data encoding, the gates, and the readout, within a small circuit language described in Appendix~\ref{app:dsl}; this tests whether the model can invent a usable architecture and program it in one shot.

\paragraph{What ``works'' means.}
Numbers written by a model are only interesting relative to what the same circuit can do. For every trial we therefore evaluate, on the same circuit layout:
\begin{itemize}
  \item \textbf{Random weights.} The same layout with every angle drawn uniformly from $[-\pi, \pi]$ (50 draws for VQE/QAOA, 20 for QML). This is the floor: it tells us how much of the score comes from the layout and the readout rather than from the numbers.
  \item \textbf{Optimizer.} A classical optimizer (L-BFGS-B with finite-difference gradients) started from random weights, with five restarts for VQE/QAOA and one for QML. For QML it sees exactly the examples the model saw, so it is the fair ``same information, numerical optimization instead of reasoning'' comparison. This is the ceiling the layout admits under modest effort, not a tuned upper bound.
  \item \textbf{Optimizer from the model's weights.} The same optimizer started where the model left off. If the model's weights are near a good basin the optimizer should reach the optimum quickly; we count the number of loss evaluations it needs to come within $0.1\%$ of the best value found from random starts.
  \item \textbf{Exact and classical references.} The exact ground energy and the brute-force maximum cut, computed by diagonalization and enumeration; and, for QML, logistic regression on the raw pixels.
\end{itemize}
For energies we report the fraction of the spectrum captured, $r = (E_{\max} - E)/(E_{\max} - E_0)$, which is $1$ at the ground state and $0$ at the highest eigenstate, and for MaxCut the usual approximation ratio $\langle C \rangle / C_{\max}$. Both put every problem on the same $[0, 1]$ scale.

\paragraph{Model.}
We use one frontier model, \gpt\ (OpenAI, Responses API), with its default reasoning settings and a 32{,}000-token output budget. Each prompt is sent 3 times (VQE, QAOA, fixed-ansatz QML) or 5 times (free-design QML) so that the spread across samples is visible; the reported mean $\pm$ standard deviation is over these repeats. Repeats are independent requests with no shared context. The harness is model-agnostic (any OpenAI-compatible or Anthropic endpoint can be substituted by a configuration entry), but this paper reports a single model.

\section{Experimental setup}
\label{sec:experiments}

Table~\ref{tab:overview} summarizes the tasks. All circuits are simulated exactly with a batched state-vector simulator written for this study and cross-checked against Qiskit on random circuits over the full gate set; the simulator, prompts, raw replies and analysis code are released (see Reproducibility).

\begin{table}[t]
\centering
\small
\begin{tabularx}{\linewidth}{lXXl}
\toprule
Family & Problem & Circuit & Model writes \\
\midrule
QML, free design & digits 3 vs 8; 4 vs 9; all 10 classes & its own (\(\leq\) 8 qubits, \(\leq\) 200 gates) & architecture + weights \\
QML, free design & ablations on 3 vs 8: 1 / 10 shots; pixels as text & its own & architecture + weights \\
QML, structured & 3 vs 8; 10 classes, with structural rules & its own (\(\geq\) 2 qubits, entangling gate) & architecture + weights \\
QML, fixed ansatz & 3 vs 8; 10 classes & amplitude encoding + HEA, \(L \in \{1, 2, 4, 8\}\) & 12, 18, 30 or 54 weights \\
VQE & transverse-field Ising chain, \(N = 6\), \(h \in \{0.5, 1, 2\}\) & HVA \(p = 2\) (4 w.); HEA \(L = 2\) (18 w.) & weights \\
VQE, scaling & critical Ising chain, \(h = 1\) & HVA \(p = 4\) (8 w.); HEA \(L = 4\) (30 w.) & weights \\
VQE & Heisenberg chain, \(N = 6\) & HEA \(L = 2\) (18 w.) & weights \\
VQE & H\(_2\) molecule, 2-qubit Hamiltonian & 1-parameter ansatz; HEA \(L = 1\) (4 w.) & weights \\
QAOA & MaxCut, \(n = 8\): three 3-regular, two Erd\H{o}s--R\'enyi, one weighted ring & standard QAOA, \(p \in \{1, 3\}\); \(p = 6\) on two graphs & \(2p\) angles \\
\bottomrule
\end{tabularx}
\caption{Tasks. Every trial is one request; the model never receives a loss, gradient or accuracy.}
\label{tab:overview}
\end{table}

\paragraph{QML on handwritten digits.}
We use the scikit-learn \texttt{digits} data set \citep{pedregosa2011scikit}: $8\times 8$ grayscale images of handwritten digits, pixel values scaled to $[0, 1]$. For each task a stratified 70/30 train/test split is fixed once; the model is shown $k$ labelled training examples per class ($k = 5$ unless stated) rendered as images (one image per class, examples side by side), and the circuit it returns is scored on the full test split (108--109 images for a pair of digits, 540 for all ten classes). The examples shown in trial $t$ are drawn with a seed fixed by $t$, so repeats differ in which examples the model sees, and conditions that share a task and $k$ see identical prompts. In the free-design setting the model writes a circuit in a small JSON language: a number of qubits, an initial state (all-zeros, or the image amplitude-encoded on six qubits), a gate list in which every angle is an affine expression of the pixels $x[i]$ with at most one pixel per angle, and a readout that maps measured bit strings to classes (Appendix~\ref{app:dsl}). The \emph{structured} condition adds three rules: at least two qubits, at least one two-qubit gate, and no two consecutive same-axis rotations on a qubit. In the fixed-ansatz setting the image is amplitude-encoded on six qubits (three qubits index the pixel row, three the column), followed by an HEA with $L \in \{1, 2, 4, 8\}$ layers; the model writes the $6(L{+}1)$ rotation angles (12 to 54) and class scores are read from the first qubit (two classes) or the first four qubits (ten classes). The deeper circuits test how far hand-written weights scale: with 54 weights the ten-class circuit is a usable model for an optimizer, as Section~\ref{sec:results} shows.

\paragraph{VQE.}
The transverse-field Ising model (TFIM) on an open chain of six spins, $H = -\sum_i Z_i Z_{i+1} - h\sum_i X_i$, is studied at $h = 0.5$ (ordered phase), $h = 1$ (critical point) and $h = 2$ (disordered phase). The Hamiltonian variational ansatz starts from $|{+}\rangle^{\otimes 6}$ and applies $p = 2$ layers of $rzz(\theta_{2k})$ on every bond followed by $rx(\theta_{2k+1})$ on every qubit; the hardware-efficient ansatz applies three layers of $ry$ rotations separated by $cx$ ladders (18 angles). At the critical point we also test a deeper HVA ($p = 4$, 8 angles) and HEA ($L = 4$, 30 angles). The antiferromagnetic Heisenberg chain $H = \sum_i (X_iX_{i+1} + Y_iY_{i+1} + Z_iZ_{i+1})$ on six spins uses the same HEA. For H$_2$ we use the two-qubit Hamiltonian of \citet{omalley2016scalable} near the equilibrium bond length (electronic energy, nuclear repulsion excluded), with a one-parameter ansatz that prepares $\cos(\theta/2)|10\rangle + \sin(\theta/2)|01\rangle$ from the Hartree--Fock state and with a four-angle HEA. The prompt lists every Pauli term of $H$ with its coefficient, the circuit gate by gate, and a one-line physical description (\eg\ ``transverse-field Ising chain of 6 spins, open boundary, $J = 1$, $h = 0.5$''); it does not give the ground energy.

\paragraph{QAOA.}
MaxCut instances on eight nodes: three random 3-regular graphs, two Erd\H{o}s--R\'enyi graphs with edge probability $0.5$, and one ring with integer edge weights in $\{1, \ldots, 5\}$. The prompt lists the edges, states the cost $C(z) = \sum_{ij} w_{ij}[z_i \neq z_j]$, writes out the circuit for $p$ layers with the convention $U_C(\gamma) = e^{-i\gamma C}$, $U_B(\beta) = e^{-i\beta\sum_i X_i}$, and asks for $(\gamma_1, \beta_1, \ldots, \gamma_p, \beta_p)$. We evaluate $p = 1$ and $p = 3$ on all six graphs and $p = 6$ (12 angles) on one 3-regular and one Erd\H{o}s--R\'enyi graph.

\section{Results}
\label{sec:results}

All 139 requests to \gpt\ produced a valid circuit or weight vector on the first attempt; the repair loop was never used. Full per-condition tables are in Appendix~\ref{app:tables}; this section walks through the three families.

\subsection{VQE: the model writes ground-state parameters}

\begin{figure}[tbp]
\centering
\includegraphics[width=0.9\linewidth]{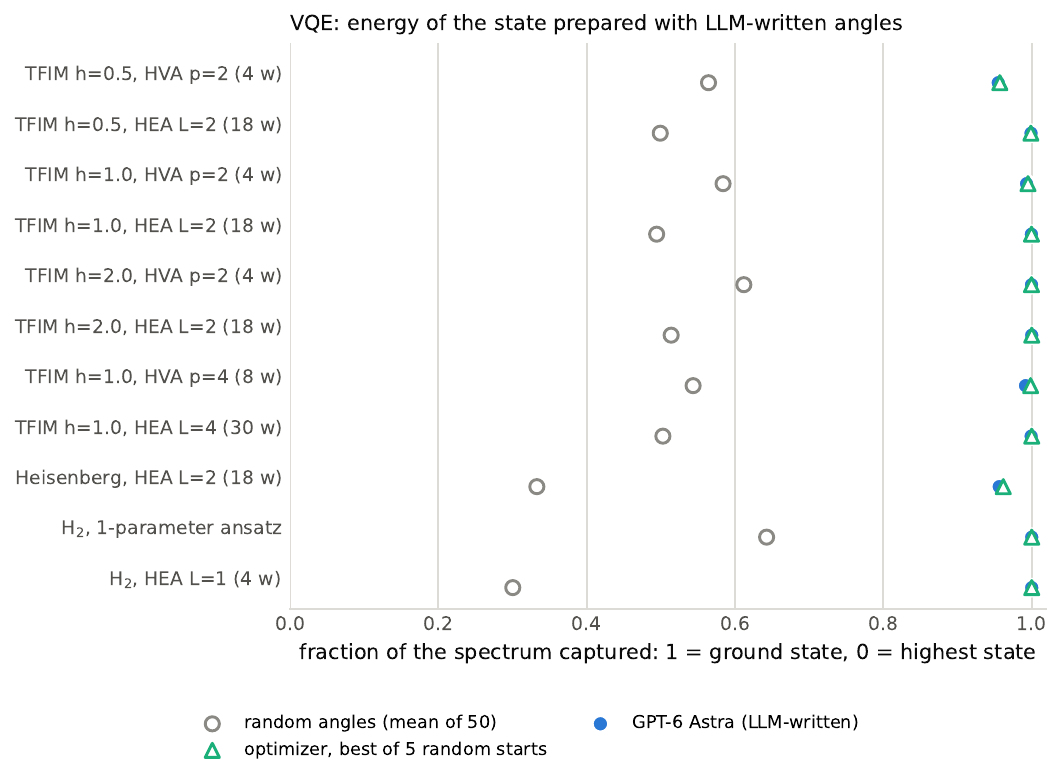}
\caption{VQE. Fraction of the spectrum captured by the state prepared with the model's angles ($1$ = exact ground state), mean $\pm$ s.d.\ over three requests, against random angles and an optimizer with five random restarts on the same circuit.}
\label{fig:vqe}
\end{figure}

Figure~\ref{fig:vqe} and Table~\ref{tab:vqe_full} summarize eleven problem--ansatz pairs. On every one the model's angles are within $0.007$ of the optimizer on the $[0, 1]$ spectrum scale, and on seven of eleven they are within $0.001$. Random angles sit at $0.30$--$0.64$.

For the transverse-field Ising chain with the Hamiltonian variational ansatz (four angles) the model reaches $0.955$, $0.993$ and $1.000$ at $h = 0.5, 1, 2$, against an optimizer ceiling of $0.957$, $0.995$ and $1.000$: the ansatz itself is the limit, and the model finds its optimum. Its angles move with the field the way the physics says they should: the coupling angle of the first layer falls from $0.55$ to $0.47$ to $0.24$ as $h$ grows from $0.5$ to $2$, while the field angle rises from $1.20$ to $1.36$, and the three repeats of each problem agree to within $0.1$ rad. With the 18-angle hardware-efficient ansatz the model reaches $0.999$--$1.000$ on all three fields; its replies describe the target state (``a ferromagnetic state with modest canting toward $+X$, strongest at the boundaries'') and set the angles to produce it. For the critical chain with a 30-angle HEA it still reaches $0.999$, by setting 19--21 of the 30 angles to exactly zero and programming one rotation layer plus one correction layer; with the eight-angle HVA it reaches $0.992$ against $0.999$ for the optimizer, the one case where more angles made the model slightly worse rather than better.

For H$_2$ the model does something different: it derives the energy as a function of the single angle, $E(\theta) = -1.0520 - 0.7782\cos\theta + 0.1820\sin\theta$ (which is correct for the stated Hamiltonian and circuit), and returns its minimizer $\theta^\ast = -\arctan(0.1820/0.7782) = -0.2297$, the exact ground state. With the four-angle HEA on the same molecule it also reaches the exact energy. The largest gaps are the eight-angle HVA ($0.007$) and the Heisenberg chain with the HEA ($0.956$ vs $0.962$); on the latter its reply reasons about singlet pairing on bonds but the real-amplitude HEA cannot represent the singlet ground state well with two layers, and both model and optimizer are limited by it.

\subsection{QAOA: the model writes angle schedules}

\begin{figure}[tbp]
\centering
\includegraphics[width=0.9\linewidth]{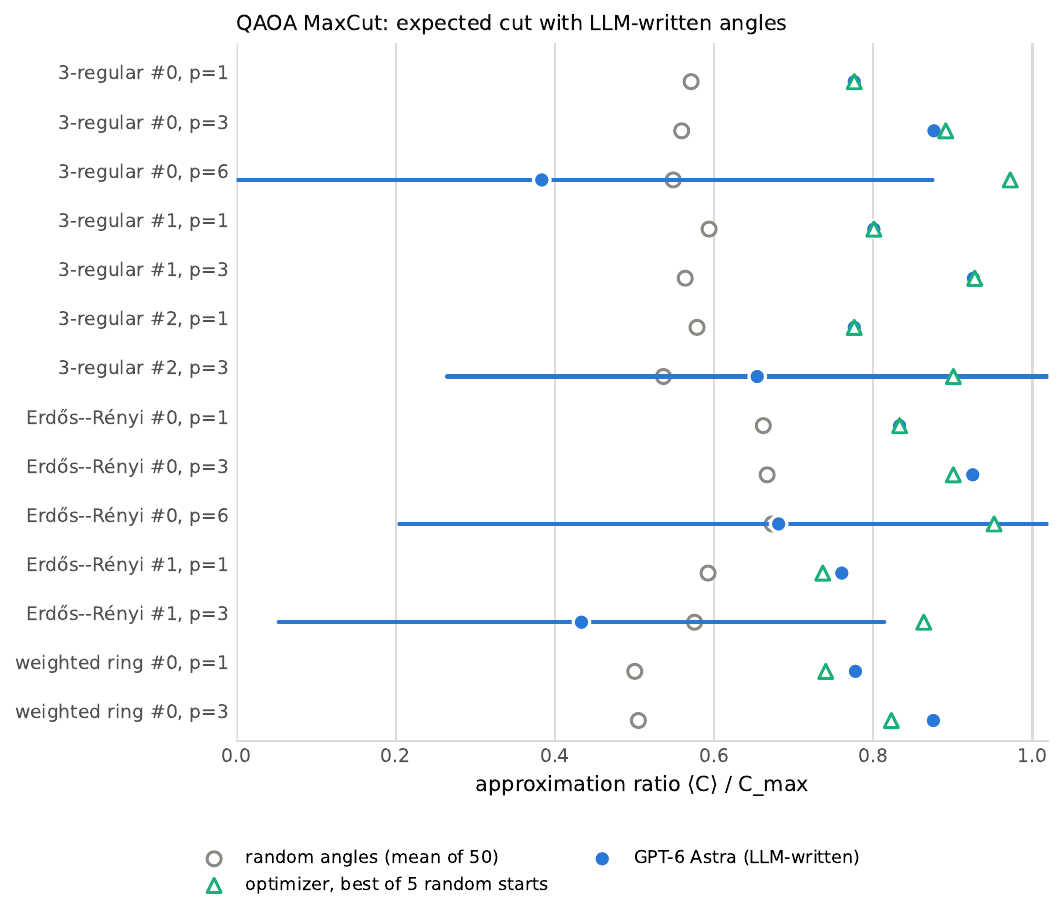}
\caption{QAOA MaxCut on eight-node graphs. Approximation ratio $\langle C\rangle / C_{\max}$ with the model's angles, against random angles and an optimizer with five random restarts. The large spreads at $p = 3$ and $p = 6$ are sign errors on the mixing angles in 6 of 42 requests (see text).}
\label{fig:qaoa}
\end{figure}

At depth $p = 1$ the model's angles are optimal on all six graphs ($0.776$, $0.801$, $0.776$, $0.833$, $0.760$, $0.778$, with zero variance across repeats): in every one of the 18 requests they reach the best value the optimizer found on that graph, and on the second Erd\H{o}s--R\'enyi graph and the weighted ring the five-restart optimizer itself fell short of that value in one run of three (mean $0.737$ and $0.740$), so the model's average is above the optimizer's. Its method is analytic: for the first 3-regular graph it writes ``the exact $p = 1$ expectation is $6 + 6\sin(4\beta)\sin(\gamma)\cos^2(\gamma) - 6\sin^2(2\beta)\sin^2(\gamma)\cos^2(\gamma)$'', accounting for the graph's four triangles, and maximizes it in closed form. At $p = 3$ and $p = 6$ it writes a smooth schedule---cost angles increasing, mixing angles decreasing across layers, a pattern known from the literature on parameter concentration---and where the schedule has the right sign it is within $0.03$ of the optimizer or above it ($0.925$ vs $0.901$ on the first Erd\H{o}s--R\'enyi graph at $p = 3$, $0.876$ vs $0.823$ on the weighted ring).

Six of the 42 requests failed in one specific way: every mixing angle was negated (\eg\ $[0.38, -0.57, 0.65, -0.47, \ldots]$ instead of $[0.36, 0.57, 0.62, 0.47, \ldots]$). Negating $\beta$ alone is not a symmetry of QAOA, and those circuits score $0.10$--$0.22$. Re-evaluating them with $|\beta|$ gives $0.87$--$0.96$, at the optimizer's level in every case. Over all 42 requests the mean ratio is $0.748$ as written and $0.856$ with the sign corrected, against $0.849$ for the optimizer; over the 36 sign-correct requests it is $0.846$ against $0.838$. We report the as-written number as the result and the corrected one as the diagnosis.

\subsection{Warm starts}

\begin{figure}[tbp]
\centering
\includegraphics[width=0.85\linewidth]{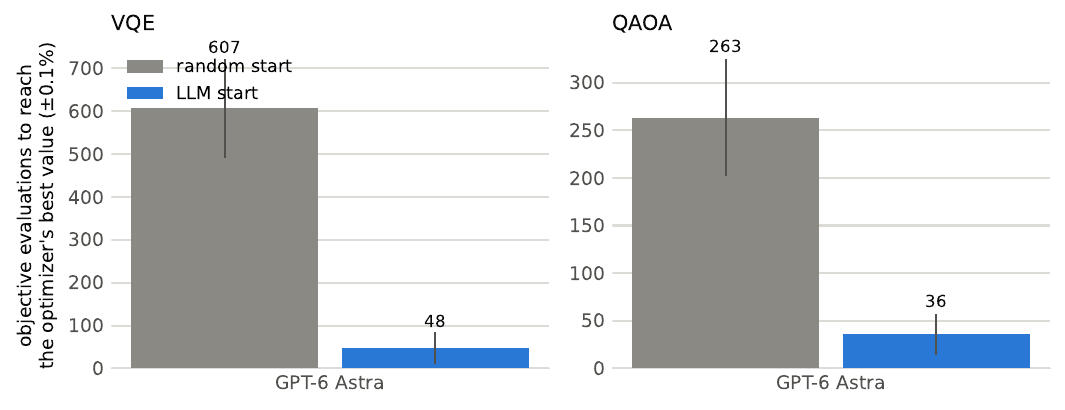}
\caption{Loss evaluations the optimizer needs to come within $0.1\%$ of the best value found from random starts, when started from random angles versus from the model's angles (mean $\pm$ s.e.\ over all VQE and QAOA requests).}
\label{fig:warmstart}
\end{figure}

Started from the model's angles, L-BFGS-B reached the target in a median of one loss evaluation in both VQE and QAOA, meaning the model's point was already within tolerance in 61\% (VQE) and 64\% (QAOA) of requests; the means are $48$ and $36$ evaluations, against $607$ and $263$ from random starts (Figure~\ref{fig:warmstart}). The warm start also succeeds more often: it reached the target in $91\%$ of VQE and $93\%$ of QAOA runs, against $81\%$ and $46\%$ for random starts. The failures to reach it from the model's angles are the sign-flipped QAOA schedules and the Heisenberg chain, where the optimizer from the model's point converges to a slightly different local minimum.

\subsection{QML: writing a classifier's weights}

\begin{figure}[tbp]
\centering
\includegraphics[width=0.9\linewidth]{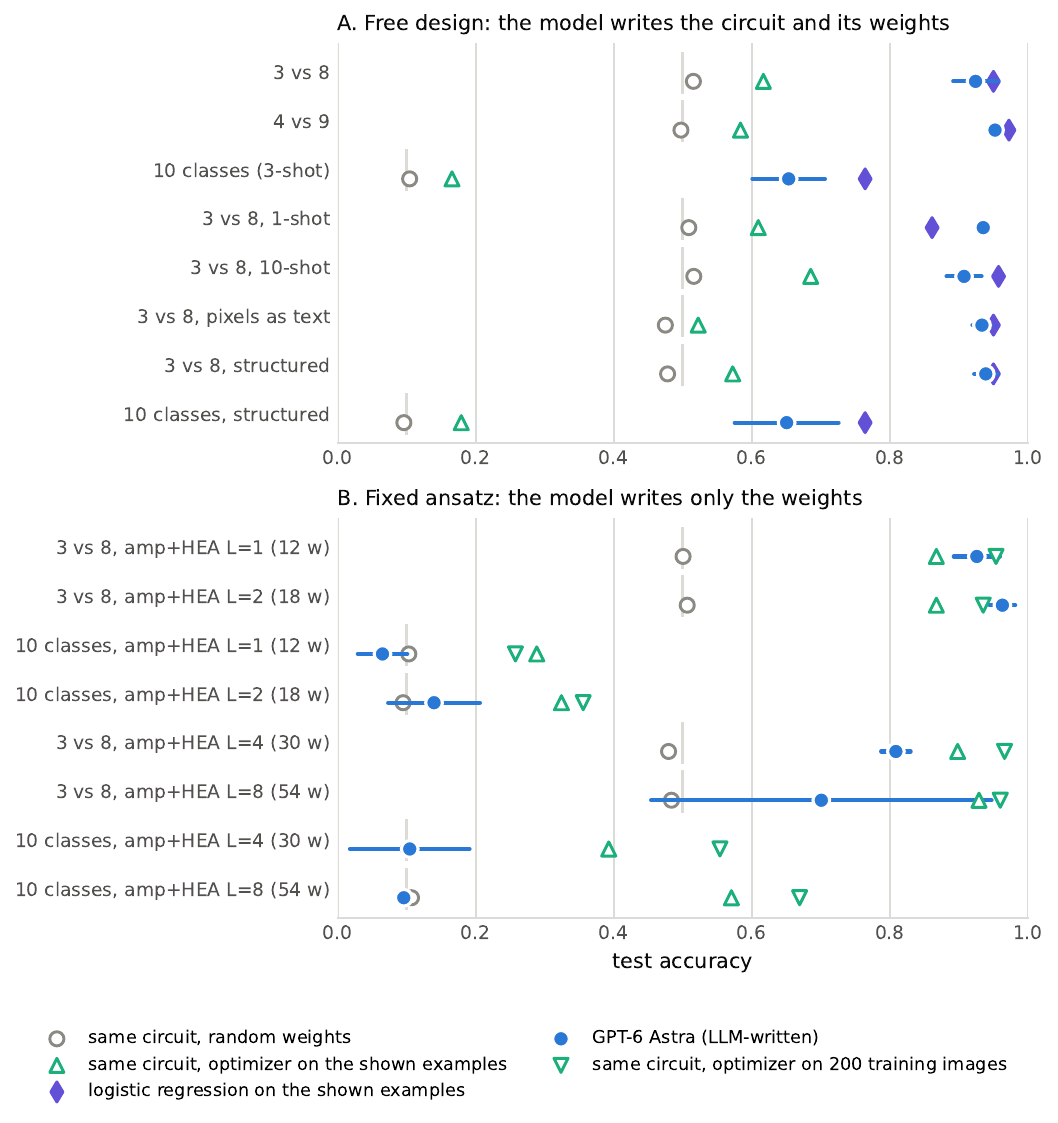}
\caption{Digit classification. Test accuracy of the model's circuit (mean $\pm$ s.d.\ over five requests in A, three in B), with the same circuit under random weights, under an optimizer that saw the same examples, and logistic regression on those examples (A) or the optimizer with 200 training images (B). Gray ticks mark chance.}
\label{fig:qml}
\end{figure}

\paragraph{Free design.}
Given five examples per class and a circuit language, the model's circuits score $0.924 \pm 0.032$ on 3 vs 8 and $0.952 \pm 0.008$ on 4 vs 9 (chance $0.5$), against $0.50$ with random weights on the same circuits and $0.58$--$0.62$ for an optimizer that trained the same circuits on the same ten examples (Figure~\ref{fig:qml}A, Table~\ref{tab:qml_free_full}). Logistic regression on the ten examples scores $0.95$--$0.97$. The number of examples barely matters: with one example per class the model scores $0.935$, with ten $0.907$, and shown the pixel grids as integers instead of images $0.933$. On all ten classes with three examples each it scores $0.653 \pm 0.053$ (chance $0.1$), against $0.17$ for the optimizer on the same thirty images and $0.76$ for logistic regression.

In all 30 unconstrained requests the model designed the same kind of circuit: chains of $ry$ rotations each driven by one pixel (plus a bias rotation), with a bit-string readout. In all 25 binary requests the circuit is a single qubit carrying one such chain of 8--51 rotations. As explained in Section~\ref{sec:discussion}, this is a linear classifier with a smooth threshold, and the model's replies describe it as one (``positive weights collect middle-left evidence, and negative weights compensate for right-side ink''). For ten classes it scaled the same idea up: six to eight qubits, each a rotation chain detecting one feature (``central versus peripheral ink, upper and lower stroke balance, crossbars, and bottom closure''), read out compositionally so that, in its words, ``5 combines a left-upper-stroke detector with lower-right dominance''. Three of the five ten-class circuits went further and used controlled-$ry$ gates, two of them with Toffoli gates as well, to implement that logic explicitly, scoring $0.67$--$0.71$. Fine-tuning the model's weights with the optimizer on the shown examples did not help on any binary task ($0.85$--$0.95$ after versus $0.91$--$0.95$ before): the optimizer, with ten examples, overfits where the model's prior does not.

\paragraph{Structured design.}
With the three structural rules (at least two qubits, at least one two-qubit gate, no consecutive same-axis rotations on a qubit) the model scores $0.939 \pm 0.015$ on 3 vs 8 and $0.650 \pm 0.067$ on ten classes, indistinguishable from the unconstrained $0.924$ and $0.653$, and all ten circuits are valid on the first attempt. What it built is the more interesting result. In every binary trial the model kept the linear accumulator and found a circuit identity that satisfies the rules: it interleaves $z$ gates between the $ry$ rotations (since $z\,ry(b)\,z = ry(-b)$, the chain still sums, with alternating signs absorbed into the weights; ``alternating Z gates and signed RY angles accumulate this evidence into one decision angle'', it explains), or accumulates the same weighted sum as a phase with $p$ or $rz$ gates between CNOTs and converts it to a probability with a final Hadamard, and then moves the result to the readout qubit with a single CNOT. For ten classes two of the five circuits are of the same kind (``the $\pi$-angle ZZ gates provide sign refocusing between successive encoding rotations''), and three use genuinely multi-qubit designs: 117 $u3$ rotations with 12 CNOTs on five qubits ($0.706$), and two eight-qubit circuits in which ``four address qubits select a class-specific affine phase; interference converts that phase into a score monotonic in a sparse nearest-prototype discriminant'', built from 39 Toffoli and 146 controlled-phase gates ($0.713$, the best ten-class circuit of the study). The model, in other words, knows exactly which weight space it wants to program---a signed sum of pixels---and, forbidden the obvious circuit for it, writes a different circuit that computes the same thing, or, when it must, a nearest-prototype classifier whose prototypes it writes by hand.

\paragraph{Fixed ansatz and scaling.}
When the circuit is fixed to amplitude encoding of the whole image followed by a hardware-efficient ansatz, the model has no interpretable pixel weights to write; it must reason about how rotations on the row and column qubits redistribute amplitude. For 3 vs 8 with 12 and 18 weights it does this well: $0.926 \pm 0.033$ and $0.963 \pm 0.019$, above the optimizer trained on the same examples ($0.867$) and close to the optimizer with 200 images ($0.935$--$0.954$). Its reply reads the circuit correctly (``3 has more ink energy in the outer row bands relative to the middle bands, while 8 is more centrally distributed; these angles implement that row-band contrast'').

\begin{figure}[tbp]
\centering
\includegraphics[width=\linewidth]{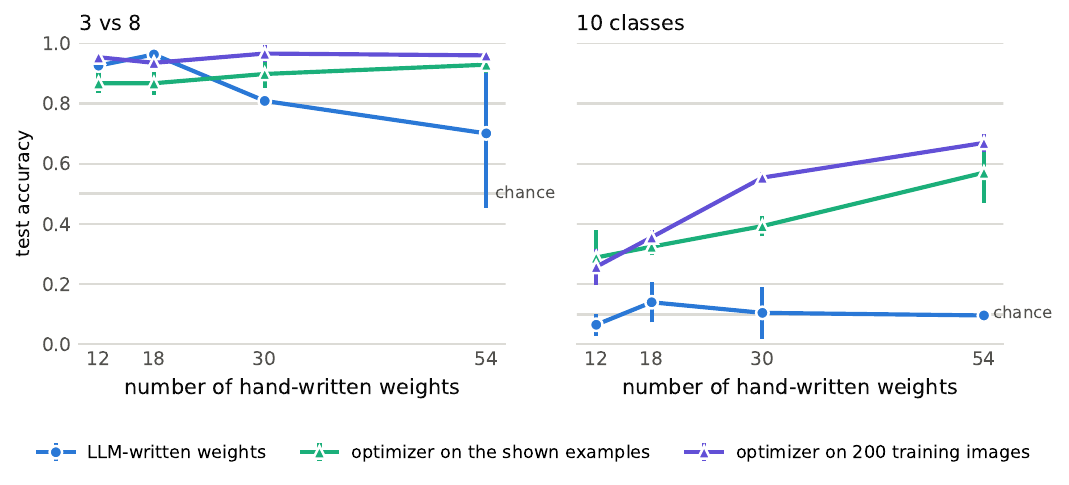}
\caption{Scaling the fixed ansatz. Test accuracy against the number of hand-written weights (amplitude encoding + HEA with 1, 2, 4 and 8 layers) for the model, the optimizer on the shown examples, and the optimizer on 200 training images.}
\label{fig:scaling}
\end{figure}

Figure~\ref{fig:scaling} shows what happens as the circuit deepens. For 3 vs 8 the optimizer's accuracy rises monotonically with depth ($0.87 \to 0.93$ on the shown examples) while the model's falls: $0.81 \pm 0.02$ at 30 weights and $0.70 \pm 0.25$ at 54. For ten classes the circuit is too weak at 12--18 weights for anyone (the optimizer on the shown examples reaches $0.29$--$0.32$, and $0.26$--$0.36$ with 200 images), but with 30 and 54 weights it becomes a usable model for the optimizer ($0.39$ and $0.57$ on the shown examples; $0.55$ and $0.67$ with 200 images) and the model stays at chance throughout ($0.07$--$0.14$). Its replies at depth are candid: ``these are hand-designed heuristic weights, not optimized or validated weights''. Started from the model's 54 weights, the optimizer reaches $0.59$, so the weights are not harmful, merely uninformative.

\subsection{Cost}
A request cost on average 1.8k output tokens and 45 s for QAOA, 3.0k tokens and 160 s for VQE, and 5.2--5.5k tokens and 100--150 s for QML; the ten-class free-design circuits were the largest replies (14k tokens, 240 s). A single evaluation of the resulting circuit on our simulator takes milliseconds; the optimizer baselines took seconds to a minute per run.

\section{Discussion}
\label{sec:discussion}

\paragraph{What the model actually does.}
Reading the replies, three strategies account for nearly all of the successes. (i)~\emph{Analytic solution.} Where the loss has a short closed form, the model derives it and solves it: for H$_2$ it writes $E(\theta) = -1.0520 - 0.7782\cos\theta + 0.1820\sin\theta$ and returns $\theta^\ast = -\arctan(0.1820/0.7782)$, which is the exact minimizer; for QAOA at $p = 1$ it writes the expectation of the cut as a trigonometric polynomial in $(\gamma, \beta)$, triangles included, and maximizes it. (ii)~\emph{Physical interpolation.} For the Ising chain it reasons about the two limits (all spins aligned at $h \to 0$, all spins along $X$ at $h \to \infty$) and places the angles between them according to $h$; its weights for the field term grow with $h$ and those for the coupling shrink. (iii)~\emph{Geometric feature design.} For digits it names regions of the image (``middle-left ink'', ``bottom closure'', ``upper loop''), assigns each region a sign and a magnitude, and sums them in a rotation. All three are forms of reasoning about what a weight does; none is search.

\paragraph{Why the free-design circuits are linear classifiers.}
Consecutive rotations about one axis add their angles, so a chain $ry(w_1 x_{i_1})\,ry(w_2 x_{i_2}) \cdots ry(b)$ on one qubit prepares a state whose probability of reading 1 is $\sin^2\!\big((b + \sum_k w_k x_{i_k})/2\big)$: a linear score passed through a smooth threshold. This is the structure the model found in every unconstrained binary trial, and one chain per qubit is what it built for ten classes; it is the natural thing to find, because it is the only circuit in the language whose weights have the meaning ``how much does this pixel count''. The structured condition, which forbids the chain, shows that the model can also program multi-qubit circuits, but it does so by finding circuit identities that compute the same weighted sum ($z$-interleaved rotations, or a phase accumulated between CNOTs and read out through a Hadamard) or, when it must, by writing a nearest-prototype classifier whose prototypes it chooses by hand. We see this as the central lesson of the free-design experiments: the model is good at programming weights whose effect it can state in a sentence, and it chooses architectures accordingly.

\paragraph{Scaling.}
The fixed-ansatz results (Figure~\ref{fig:scaling}) show the limit cleanly. The amplitude-encoded circuit is a fixed unitary applied to the image vector, so each class score is the squared projection of the image onto one direction, and the weights control that direction only through a product of many rotations. At 12--18 weights the model can still say what the circuit computes (``row-band contrast'', in its words) and its weights beat the optimizer; at 30--54 weights it cannot, it says so (``hand-designed heuristic weights''), and its accuracy decays toward chance while the optimizer's improves. For ten classes the circuit needs at least 54 weights before an optimizer makes it useful, and the model is at chance throughout. Weight programming therefore scales with the \emph{interpretability} of the weight space rather than with its size: the HVA has only four to eight weights, each with a physical meaning, and the model handles them well; the 30-weight HEA for the critical Ising chain is also handled (ratio 0.999), but there the model treats most weights as zero and programs a few. Problems whose solution needs many weights with no individual meaning are beyond the method, and the model knows it. The contrast with \citet{takehi2026sparks}, where the same model wrote networks of $10^4$--$10^6$ weights, is instructive: those weights live in architectures the model can decompose (a convolution that detects a paddle, a layer that reads a board), so a million numbers can be written as a few hundred decisions. Fifty-four angles whose only effect is through a product of rotations on a $64$-dimensional amplitude vector offer no such decomposition, and are harder to program than a million structured ones.

\paragraph{Warm starts.}
Even where the model's weights are not optimal, they are a good place to start. Across all VQE and QAOA trials the optimizer started from the model's angles reached its target after a median of one loss evaluation, that is, the model's point was already within tolerance, and in the remaining cases after tens rather than hundreds of evaluations. Since loss evaluations are runs of the quantum device, this is the most direct practical consequence of the results: a few seconds of model reasoning replaces most of the optimization loop.

\paragraph{Conventions and failure modes.}
The one systematic failure is a sign. In six QAOA trials every mixing angle came out negated; the magnitudes were right and the sign-corrected circuits match the optimizer. The prompt stated the convention $U_B(\beta) = e^{-i\beta\sum X}$ explicitly, and the gate list showed $rx(2\theta)$, so this is a reasoning slip rather than missing information, but it illustrates that weight programming is brittle in a way optimization is not: an optimizer does not care which direction is positive. In practice a single loss evaluation would catch it; we chose not to allow one, to keep the test clean.

\paragraph{Limitations.}
All circuits are small (at most eight qubits) and simulated without noise. The problems are well known, and the model has certainly seen the Ising model, QAOA on regular graphs and the digits data set during training; part of what we measure is recall. We regard this as a feature of the method rather than a confound (a programmer's prior knowledge is the point), but it means the results do not predict performance on novel Hamiltonians or data sets. The optimizer baseline is deliberately modest (L-BFGS-B with finite differences, a few restarts, a 60-second budget) and is a reference, not a tuned upper bound; this matters most for the deep fixed-ansatz QML cases, where a stronger optimizer would widen the gap to the model. We report a single model; whether other frontier models share its strategies and its failure modes is an open question that the released harness is built to answer. Finally, our scoring never reaches the model, which is the strictest possible setting; allowing even one round of feedback would likely remove the sign failures and improve the deep circuits, and is the obvious next experiment.

\section{Related work}
\label{sec:related}

\paragraph{Weight-space programming.}
The term and the idea are due to \citet{takehi2026sparks}, who showed with the same model we use that a frontier LLM can read the raw weights of small networks and name the game or data they were trained on, edit $3.5\%$ of a failing control policy's weights to make it succeed, and write from scratch, with no training and no environment rollouts, the weights of MinAtar game policies that are competitive with or better than PPO-trained ones and of a $9{\times}9$ Go player that beats smaller PPO networks; they also found that weight editing removes obstacles (a looping behaviour in a small language model) but does not add missing abilities. Our study transfers the question to variational quantum circuits. The differences in regime are what make the transfer interesting: our networks have 1--54 weights rather than thousands to millions, the weights are angles with a physical or geometric meaning, the loss is an exact expectation value rather than a game score, and the optimizer being replaced is one whose every evaluation runs a quantum device.

\paragraph{Initialization and warm starts for VQAs.}
Because VQA training is hard in general \citep{bittel2021training} and random initializations land on barren plateaus \citep{mcclean2018barren}, a line of work seeks good starting points: identity-block initializations \citep{grant2019initialization}, warm starts from relaxed classical solutions \citep{egger2021warm}, and for QAOA the observation that optimal angles concentrate across instances \citep{brandao2018fixed,zhou2020quantum} and can be tabulated for regular graphs \citep{wurtz2021fixed}. Weight-space programming can be read as initialization by a reasoning system that has absorbed this literature; our warm-start measurements quantify how much optimizer effort it saves.

\paragraph{Language models as optimizers and program synthesizers.}
\citet{yang2024large} showed that a language model can act as an optimizer when it is shown a trajectory of candidate solutions and their scores (OPRO), and FunSearch \citep{romera2024mathematical} evolves programs under a scorer. Both are feedback loops. Our setting removes the loop: the model answers once, without any score, so what is measured is prior knowledge and reasoning rather than in-context optimization.

\paragraph{Generating weights instead of training them.}
Hypernetworks \citep{ha2017hypernetworks} generate the weights of one network with another, trained end to end. Our models are not trained for the task, and the ``hypernetwork'' is a general-purpose language model producing weights as text.

\paragraph{Expressibility and encodings.}
The capacity of a PQC depends on its layout \citep{sim2019expressibility} and on how data enters it \citep{schuld2021effect,havlivcek2019supervised}. The free-design results above illustrate this from an unusual angle: left to choose, the model picks the simplest encoding that makes its weights interpretable, which is also the one with the least quantum structure.

\section{Conclusion}
\label{sec:conclusion}

A frontier language model can write the parameters of a variational quantum circuit directly, with no optimizer and no feedback, whenever the parameters have a meaning it can reason about: the angles of a Hamiltonian variational ansatz, the schedule of a QAOA circuit, the pixel weights of a small classifier. In those settings its weights match a classical optimizer's, sometimes exceed an optimizer given the same data, and almost always put an optimizer within one evaluation of its best value. The approach does not extend to circuits whose weights have no individual meaning, and it is vulnerable to convention errors that an optimizer would never make. We think the right reading, in line with \citet{takehi2026sparks}, is that language models are not optimizers but programmers, and that variational quantum circuits are a natural home for weight-space programming: their weights are few, meaningful, and expensive to train.

\section*{Reproducibility}
All code, configuration files, prompts, raw model replies, parsed circuits, per-trial scores and the analysis scripts that produce every figure and table in this paper are available at \url{https://github.com/DebarshiKunduPSU/Quantum-Weight-Space-Programming}. Each trial directory holds the exact prompt (with the images shown to the model), every reply including repair turns, the circuit that was simulated, and a JSON record of all scores and baselines. \texttt{scripts/run\_all.sh <model>} reruns every experiment for a model, skipping finished trials; \texttt{python -m weightprog.analysis} regenerates the figures and tables. The simulator is verified against Qiskit by \texttt{pytest}. Model replies are stochastic, so a rerun will not reproduce our numbers exactly; the released traces are the record of the runs reported here.

\bibliographystyle{plainnat}
\bibliography{refs}

\appendix
\section{The circuit language for free design}
\label{app:dsl}

In the free-design QML experiments the model writes a JSON object with four fields. \texttt{n\_qubits} is the register size. \texttt{state\_prep} is either \texttt{\{"type": "zero"\}} or \texttt{\{"type": "amplitude", "qubits": [q0, \ldots, q5]\}}, which loads the normalized 64-pixel vector as the amplitudes of the six listed qubits (the first three index the row, the last three the column). \texttt{gates} is an ordered list; each entry names a gate from \{h, x, y, z, s, sdg, t, tdg, sx, rx, ry, rz, p, u3, cx, cy, cz, swap, crx, cry, crz, cp, rxx, ryy, rzz, ccx\}, its qubits, and for rotations an angle written as a number or as a string expression over numbers, \texttt{pi}, pixels \texttt{x[i]}, and $+\,-\,*\,/$. Angles must be affine in the pixels and, in the default condition, contain at most one pixel. \texttt{readout} is either \texttt{bitstrings} (a list of measured qubits and, per class, the bit strings whose total probability is the class score) or \texttt{qubit\_per\_class} (the class score is the probability that its qubit reads 1). The predicted class is the one with the highest score. The complete system prompt, including these rules, is in the repository (\texttt{weightprog/prompts.py}) and in every trial directory as \texttt{prompt.txt}.

\section{Full result tables}
\label{app:tables}

Tables~\ref{tab:qml_free_full}--\ref{tab:qaoa_full} list every condition: validity (usable circuits / requests), the score of the model-written weights, the random-weight floor, the optimizer from random starts, and the optimizer started from the model's weights. Means $\pm$ standard deviations are over repeated requests.

\begin{table}[h]
\centering\footnotesize
\resizebox{\linewidth}{!}{\begin{tabular}{llcccccc}
\toprule
Task & Model & Valid & LLM & Random w. & Optimizer & LLM$\to$opt. & LogReg \\
\midrule
3 vs 8 & GPT-6 Astra & 5/5 & 0.924 $\pm$ 0.032 & 0.515 $\pm$ 0.018 & 0.617 $\pm$ 0.094 & 0.869 $\pm$ 0.094 & 0.950 $\pm$ 0.024 \\
4 vs 9 & GPT-6 Astra & 5/5 & 0.952 $\pm$ 0.008 & 0.497 $\pm$ 0.013 & 0.583 $\pm$ 0.097 & 0.950 $\pm$ 0.026 & 0.972 $\pm$ 0.015 \\
10 classes (3-shot) & GPT-6 Astra & 5/5 & 0.653 $\pm$ 0.053 & 0.104 $\pm$ 0.007 & 0.166 $\pm$ 0.082 & 0.636 $\pm$ 0.045 & 0.764 $\pm$ 0.016 \\
3 vs 8, 1-shot & GPT-6 Astra & 5/5 & 0.935 $\pm$ 0.007 & 0.509 $\pm$ 0.009 & 0.609 $\pm$ 0.107 & 0.926 $\pm$ 0.025 & 0.861 $\pm$ 0.066 \\
3 vs 8, 10-shot & GPT-6 Astra & 5/5 & 0.907 $\pm$ 0.025 & 0.516 $\pm$ 0.040 & 0.685 $\pm$ 0.164 & 0.846 $\pm$ 0.057 & 0.957 $\pm$ 0.011 \\
3 vs 8, pixels as text & GPT-6 Astra & 5/5 & 0.933 $\pm$ 0.014 & 0.475 $\pm$ 0.028 & 0.522 $\pm$ 0.063 & 0.863 $\pm$ 0.081 & 0.950 $\pm$ 0.024 \\
3 vs 8, structured & GPT-6 Astra & 5/5 & 0.939 $\pm$ 0.017 & 0.478 $\pm$ 0.014 & 0.572 $\pm$ 0.060 & 0.898 $\pm$ 0.048 & 0.950 $\pm$ 0.024 \\
10 classes, structured & GPT-6 Astra & 5/5 & 0.650 $\pm$ 0.075 & 0.096 $\pm$ 0.004 & 0.179 $\pm$ 0.077 & 0.529 $\pm$ 0.102 & 0.764 $\pm$ 0.016 \\
\bottomrule
\end{tabular}
}
\caption{Free-design QML, test accuracy.}
\label{tab:qml_free_full}
\end{table}

\begin{table}[h]
\centering\footnotesize
\resizebox{\linewidth}{!}{\begin{tabular}{llcccccc}
\toprule
Task, ansatz & Model & Valid & LLM & Random w. & Opt. (shown) & LLM$\to$opt. & Opt. (200 img.) \\
\midrule
3 vs 8, amp+HEA L=1 (12 w) & GPT-6 Astra & 3/3 & 0.926 $\pm$ 0.033 & 0.500 $\pm$ 0.023 & 0.867 $\pm$ 0.033 & 0.867 $\pm$ 0.033 & 0.954 $\pm$ 0.000 \\
3 vs 8, amp+HEA L=2 (18 w) & GPT-6 Astra & 3/3 & 0.963 $\pm$ 0.019 & 0.506 $\pm$ 0.018 & 0.867 $\pm$ 0.039 & 0.867 $\pm$ 0.039 & 0.935 $\pm$ 0.000 \\
10 classes, amp+HEA L=1 (12 w) & GPT-6 Astra & 3/3 & 0.065 $\pm$ 0.036 & 0.103 $\pm$ 0.009 & 0.288 $\pm$ 0.092 & 0.243 $\pm$ 0.076 & 0.257 $\pm$ 0.059 \\
10 classes, amp+HEA L=2 (18 w) & GPT-6 Astra & 3/3 & 0.140 $\pm$ 0.066 & 0.095 $\pm$ 0.008 & 0.324 $\pm$ 0.029 & 0.246 $\pm$ 0.023 & 0.356 $\pm$ 0.025 \\
3 vs 8, amp+HEA L=4 (30 w) & GPT-6 Astra & 3/3 & 0.809 $\pm$ 0.021 & 0.479 $\pm$ 0.054 & 0.898 $\pm$ 0.046 & 0.920 $\pm$ 0.075 & 0.966 $\pm$ 0.028 \\
3 vs 8, amp+HEA L=8 (54 w) & GPT-6 Astra & 3/3 & 0.701 $\pm$ 0.247 & 0.483 $\pm$ 0.049 & 0.929 $\pm$ 0.028 & 0.938 $\pm$ 0.023 & 0.960 $\pm$ 0.005 \\
10 classes, amp+HEA L=4 (30 w) & GPT-6 Astra & 3/3 & 0.104 $\pm$ 0.087 & 0.102 $\pm$ 0.004 & 0.393 $\pm$ 0.034 & 0.408 $\pm$ 0.044 & 0.554 $\pm$ 0.020 \\
10 classes, amp+HEA L=8 (54 w) & GPT-6 Astra & 3/3 & 0.096 $\pm$ 0.006 & 0.107 $\pm$ 0.006 & 0.570 $\pm$ 0.100 & 0.594 $\pm$ 0.024 & 0.669 $\pm$ 0.029 \\
\bottomrule
\end{tabular}
}
\caption{Fixed-ansatz QML, test accuracy.}
\label{tab:qml_fixed_full}
\end{table}

\begin{table}[h]
\centering\footnotesize
\resizebox{\linewidth}{!}{\begin{tabular}{llccccc}
\toprule
Problem, ansatz & Model & Valid & LLM & Random & Optimizer & LLM$\to$opt. \\
\midrule
TFIM h=0.5, HVA p=2 (4 w) & GPT-6 Astra & 3/3 & 0.955 $\pm$ 0.001 & 0.564 $\pm$ 0.026 & 0.957 $\pm$ 0.000 & 0.957 $\pm$ 0.000 \\
TFIM h=0.5, HEA L=2 (18 w) & GPT-6 Astra & 3/3 & 0.999 $\pm$ 0.001 & 0.499 $\pm$ 0.026 & 0.999 $\pm$ 0.000 & 0.999 $\pm$ 0.001 \\
TFIM h=1.0, HVA p=2 (4 w) & GPT-6 Astra & 3/3 & 0.993 $\pm$ 0.001 & 0.584 $\pm$ 0.004 & 0.995 $\pm$ 0.000 & 0.995 $\pm$ 0.000 \\
TFIM h=1.0, HEA L=2 (18 w) & GPT-6 Astra & 3/3 & 0.999 $\pm$ 0.000 & 0.494 $\pm$ 0.012 & 1.000 $\pm$ 0.000 & 1.000 $\pm$ 0.000 \\
TFIM h=2.0, HVA p=2 (4 w) & GPT-6 Astra & 3/3 & 1.000 $\pm$ 0.000 & 0.612 $\pm$ 0.014 & 1.000 $\pm$ 0.000 & 1.000 $\pm$ 0.000 \\
TFIM h=2.0, HEA L=2 (18 w) & GPT-6 Astra & 3/3 & 1.000 $\pm$ 0.000 & 0.514 $\pm$ 0.008 & 1.000 $\pm$ 0.000 & 1.000 $\pm$ 0.000 \\
TFIM h=1.0, HVA p=4 (8 w) & GPT-6 Astra & 3/3 & 0.992 $\pm$ 0.003 & 0.543 $\pm$ 0.019 & 0.999 $\pm$ 0.000 & 0.999 $\pm$ 0.000 \\
TFIM h=1.0, HEA L=4 (30 w) & GPT-6 Astra & 3/3 & 0.999 $\pm$ 0.001 & 0.503 $\pm$ 0.011 & 1.000 $\pm$ 0.000 & 1.000 $\pm$ 0.000 \\
Heisenberg, HEA L=2 (18 w) & GPT-6 Astra & 3/3 & 0.956 $\pm$ 0.005 & 0.333 $\pm$ 0.016 & 0.962 $\pm$ 0.000 & 0.959 $\pm$ 0.000 \\
H$_2$, 1-parameter ansatz & GPT-6 Astra & 3/3 & 1.000 $\pm$ 0.000 & 0.643 $\pm$ 0.033 & 1.000 $\pm$ 0.000 & 1.000 $\pm$ 0.000 \\
H$_2$, HEA L=1 (4 w) & GPT-6 Astra & 3/3 & 1.000 $\pm$ 0.000 & 0.300 $\pm$ 0.034 & 1.000 $\pm$ 0.000 & 1.000 $\pm$ 0.000 \\
\bottomrule
\end{tabular}
}
\caption{VQE, fraction of the spectrum captured ($1$ = exact ground state).}
\label{tab:vqe_full}
\end{table}

\begin{table}[h]
\centering\footnotesize
\resizebox{\linewidth}{!}{\begin{tabular}{llccccc}
\toprule
Graph, depth & Model & Valid & LLM & Random & Optimizer & LLM$\to$opt. \\
\midrule
3-regular \#0, p=1 & GPT-6 Astra & 3/3 & 0.776 $\pm$ 0.000 & 0.571 $\pm$ 0.020 & 0.776 $\pm$ 0.000 & 0.776 $\pm$ 0.000 \\
3-regular \#0, p=3 & GPT-6 Astra & 3/3 & 0.876 $\pm$ 0.005 & 0.559 $\pm$ 0.015 & 0.891 $\pm$ 0.016 & 0.901 $\pm$ 0.000 \\
3-regular \#0, p=6 & GPT-6 Astra & 3/3 & 0.383 $\pm$ 0.491 & 0.549 $\pm$ 0.018 & 0.972 $\pm$ 0.010 & 0.979 $\pm$ 0.012 \\
3-regular \#1, p=1 & GPT-6 Astra & 3/3 & 0.801 $\pm$ 0.000 & 0.594 $\pm$ 0.006 & 0.801 $\pm$ 0.000 & 0.801 $\pm$ 0.000 \\
3-regular \#1, p=3 & GPT-6 Astra & 3/3 & 0.926 $\pm$ 0.004 & 0.564 $\pm$ 0.014 & 0.927 $\pm$ 0.015 & 0.936 $\pm$ 0.000 \\
3-regular \#2, p=1 & GPT-6 Astra & 3/3 & 0.776 $\pm$ 0.000 & 0.578 $\pm$ 0.014 & 0.776 $\pm$ 0.000 & 0.776 $\pm$ 0.000 \\
3-regular \#2, p=3 & GPT-6 Astra & 3/3 & 0.654 $\pm$ 0.390 & 0.536 $\pm$ 0.008 & 0.901 $\pm$ 0.000 & 0.891 $\pm$ 0.016 \\
Erd\H{o}s--R\'enyi \#0, p=1 & GPT-6 Astra & 3/3 & 0.833 $\pm$ 0.000 & 0.662 $\pm$ 0.007 & 0.833 $\pm$ 0.000 & 0.833 $\pm$ 0.000 \\
Erd\H{o}s--R\'enyi \#0, p=3 & GPT-6 Astra & 3/3 & 0.925 $\pm$ 0.002 & 0.667 $\pm$ 0.005 & 0.901 $\pm$ 0.032 & 0.934 $\pm$ 0.000 \\
Erd\H{o}s--R\'enyi \#0, p=6 & GPT-6 Astra & 3/3 & 0.681 $\pm$ 0.477 & 0.673 $\pm$ 0.003 & 0.952 $\pm$ 0.006 & 0.965 $\pm$ 0.023 \\
Erd\H{o}s--R\'enyi \#1, p=1 & GPT-6 Astra & 3/3 & 0.760 $\pm$ 0.000 & 0.592 $\pm$ 0.014 & 0.737 $\pm$ 0.041 & 0.760 $\pm$ 0.000 \\
Erd\H{o}s--R\'enyi \#1, p=3 & GPT-6 Astra & 3/3 & 0.433 $\pm$ 0.380 & 0.575 $\pm$ 0.014 & 0.863 $\pm$ 0.033 & 0.882 $\pm$ 0.000 \\
weighted ring \#0, p=1 & GPT-6 Astra & 3/3 & 0.778 $\pm$ 0.000 & 0.500 $\pm$ 0.011 & 0.740 $\pm$ 0.065 & 0.778 $\pm$ 0.000 \\
weighted ring \#0, p=3 & GPT-6 Astra & 3/3 & 0.876 $\pm$ 0.004 & 0.505 $\pm$ 0.024 & 0.823 $\pm$ 0.019 & 0.893 $\pm$ 0.000 \\
\bottomrule
\end{tabular}
}
\caption{QAOA MaxCut, approximation ratio.}
\label{tab:qaoa_full}
\end{table}

\end{document}